\documentclass[%
 reprint,
superscriptaddress,
 amsmath,amssymb,
 aps,
]{revtex4-2}
\usepackage{hyperref}
\usepackage{graphicx}
\usepackage{dcolumn}
\usepackage{bm}
\usepackage{siunitx}

\begin{document}

\preprint{APS/123-QED}

\title{Observing the suppression of superconductivity in \texorpdfstring{{RbEuFe\textsubscript{4}As\textsubscript{4}}{}}~ by correlated magnetic fluctuations}

\author{D. Collomb}\email{d.collomb@bath.ac.uk}
\affiliation{University of Bath, Claverton Down, Bath, BA2 7AY, United Kingdom
} 
 \author{S. J. Bending}
\affiliation{University of Bath, Claverton Down, Bath, BA2 7AY, United Kingdom
} 
\author{A. E. Koshelev}
\affiliation{
 Materials Science Division, Argonne National Laboratory, 9700 S. Cass Ave., Lemont, Illinois 60439, USA
}
 \author{M. P. Smylie}
  \affiliation{
 Materials Science Division, Argonne National Laboratory, 9700 S. Cass Ave., Lemont, Illinois 60439, USA
}
 \affiliation{Department of Physics and Astronomy, Hofstra University, Hempstead, New York, 11549, USA}
\author{L. Farrar}
\affiliation{University of Bath, Claverton Down, Bath, BA2 7AY, United Kingdom
} 
\author{J.-K. Bao}
\affiliation{
 Materials Science Division, Argonne National Laboratory, 9700 S. Cass Ave., Lemont, Illinois 60439, USA
}
\affiliation{
Laboratory of Crystallography, University of Bayreuth, D-95447 Bayreuth
Germany
}
\author{D. Y. Chung}
\affiliation{
 Materials Science Division, Argonne National Laboratory, 9700 S. Cass Ave., Lemont, Illinois 60439, USA
}
\author{M. G. Kanatzidis}
\affiliation{
 Materials Science Division, Argonne National Laboratory, 9700 S. Cass Ave., Lemont, Illinois 60439, USA
}
\affiliation{
 Department of Chemistry, Northwestern University, Evanston, Illinois, 60208, USA
}
\author{W.-K. Kwok}
\author{U. Welp}
\affiliation{
 Materials Science Division, Argonne National Laboratory, 9700 S. Cass Ave., Lemont, Illinois 60439, USA
}
\date{\today}

\begin{abstract}
In this letter, we describe quantitative magnetic imaging of superconducting vortices in RbEuFe\textsubscript{4}As\textsubscript{4} in order to investigate the unique interplay between the magnetic and superconducting sublattices. Our scanning Hall microscopy data reveal a pronounced suppression of the superfluid density near the magnetic ordering temperature in good qualitative agreement with a recently-developed model describing the suppression of superconductivity by correlated magnetic fluctuations. These results indicate a pronounced exchange interaction between the superconducting and magnetic subsystems in RbEuFe\textsubscript{4}As\textsubscript{4} with important implications for future investigations of physical phenomena arising from the interplay between them.

\end{abstract}

\maketitle

The interplay between magnetism and superconductivity has intrigued scientists for decades \cite{FuldeKellerInBook1982,Bulaevskii85Coexistence,Kulic2008InBook}. Unlike the coexistence of ferromagnetism and superconductivity in unconventional spin-triplet uranium compounds \cite{saxena2000superconductivity,Aoki2019UReview}, their coexistence in spin-singlet superconductors is generally unfavourable because the magnetic exchange field destroys opposite spin Cooper pairs
\cite{FuldeKellerInBook1982,Bulaevskii85Coexistence,Kulic2008InBook}.
Nevertheless, a growing number of rare spin-singlet superconductors with a magnetic transition temperature, T\textsubscript{m}, \textit{below} the superconducting transition temperature, T\textsubscript{c},  has been discovered. This includes the rare-earth (R) based materials RRh\textsubscript{4}B\textsubscript{4} \cite{maple1982superconductivity}, RMo\textsubscript{8}S\textsubscript{8} \cite{maple1982magnetic}, in which magnetic ordering eventually destroys superconductivity, and also the nickel borocarbides with full co-existence of superconductivity and magnetism \cite{muller2001interaction, wulferding2015spatially}. 
The magnetic moments in these compounds 
reside in sublattices that are spatially separated from the superconducting electrons, thus the magnetic exchange interaction is weak enough to allow for the coexistence of superconductivity and magnetism below their respective transition temperatures \cite{eisaki1994competition}.


One family with growing prominence in this field is the europium-containing 
iron pnictides \cite{cao2012coexistence,Zapf2017EuReview}. These typically exhibit high T\textsubscript{c}s in excess of 30K, and somewhat lower magnetic ordering temperatures (15K-20K). Hence the strong superconducting pairing, relatively large magnetic exchange interaction and wide temperature window makes them ideal materials to investigate emerging new physical phenomena.
Unlike the Eu-122 compounds, which require doping \cite{ren2009superconductivity, jeevan2008high, maiwald2012signatures, paulose2011doping, qi2012superconductivity}, or the application of pressure to obtain superconducting and magnetic transitions \cite{miclea2009evidence, matsubayashi2011pressure}, the stoichiometric 1144 compounds (e.g.\ RbEuFe\textsubscript{4}As\textsubscript{4} and CsEuFe\textsubscript{4}As\textsubscript{4}) yield both under ambient conditions
\cite{Liu2016Superconductivity,Kawashima2016Superconductivity,bao2018single, liu2016new}. The Eu atoms in RbEuFe\textsubscript{4}As\textsubscript{4} carry large, spin-only moments that undergo long-range ordering at 15K. Below the magnetic transition temperature these moments exhibit in-plane alignment, and there is a large anisotropy of the in-plane and out-of-plane exchange constants \cite{smylie2018anisotropic}. This makes it distinct from materials where the moments order along the c-axis, which can create their own unique states of vortex matter linked to ferromagnetic stripe domain structures such as in EuFe\textsubscript{2}(As\textsubscript{0.79}P\textsubscript{0.21})\textsubscript{2} \cite{stolyarov2018domain}. 
Neutron scattering experiments on RbEuFe\textsubscript{4}As\textsubscript{4} have revealed helical ordering of successive layers with a period of 4 unit cells along the c-axis due to a weak antiferromagnetic exchange interaction in this direction \cite{iida2019coexisting}.

Although the magnetic structure in RbEuFe\textsubscript{4}As\textsubscript{4} is now
quite well understood, its impact on the coexisting superconductivity is still unclear. Above the magnetic ordering temperature fluctuating magnetic moments are thought to suppress superconductivity via magnetic scattering\cite{AbrGorJETP61,SkalskiPhysRev.136.A1500,KoganLonPenDepthPhysRevB13}, while in the vicinity of T\textsubscript{m} these moments become strongly correlated, further enhancing this suppression\cite{RainerZPhys1972,MachidaJLTP1979,koshelev2020suppression}. Optical conductivity measurements probing the RbEuFe\textsubscript{4}As\textsubscript{4} superconducting gap revealed a small drop in $\Delta(T)$ as T\textsubscript{m} is approached from above, followed by a recovery at lower temperatures \cite{stolyarov2018unique}. Additionally, magnetic force microscopy (MFM) imaging of vortices revealed a gradual reduction in vortex density below $\sim$18K, which drops to a weak minimum at $\sim$12K and recovers again at lower temperatures, further hinting at a weak interaction between the superconducting and magnetic subsystems \cite{stolyarov2018unique}.
The analysis of the MFM measurements was, however, limited to counting vortex numbers as a function of temperature, rather than a direct investigation of the vortex structures themselves. On the other hand, recent angle resolved photoemission spectroscopy (ARPES) reveals no significant suppression of the superconducting gaps around the magnetic ordering temperature and DFT calculations show that the topology and orbital character of the Fe3d bands do not strongly depend on the magnetic order, although the band structure evidently exhibits a degree of sensitivity to it \cite{kim2020superconductivity}. However, these measurements would not capture the full impact of the effect of magnetic fluctuations on superconductivity, as the relative exchange correction to the gap is predicted to be significantly smaller than the correction to the superfluid density \cite{koshelev2020suppression}.
%
Here we use high-resolution scanning Hall microscopy (SHM) to investigate the influence of magnetism on individual superconducting vortices and directly extract the temperature dependence of the penetration depth, $\lambda(T)$, and the superfluid density, $\rho_{s}(T)$. This approach has the advantage that it is not influenced by the statistical nature of vortex patterns or by internal flux pumping effects by the magnetic sublattice \cite{vlasko2019self}.

The exchange interaction between the localized moments and Cooper pairs is expected to suppress the superfluid density. In the paramagnetic phase, in the regime when the exchange-field correlation length $\xi_h$ is much smaller than the in-plane coherence length $\xi_s$, this suppression is caused by magnetic scattering and very similar to the case of magnetic impurities \cite{SkalskiPhysRev.136.A1500,KoganLonPenDepthPhysRevB13}. However, as the correlation length diverges for $T\rightarrow T_{m}$, it always exceeds $\xi_s$ in the vicinity of $T_{m}$ leading to a different 'smooth' regime of interaction between the magnetic and superconducting subsystems. In the case of RbEuFe\textsubscript{4}As\textsubscript{4}, $\xi_s$ is very small \cite{smylie2018anisotropic, willa2019strongly}, resulting in a 'smooth' regime across a significant temperature range. The crossover between these 'scattering' and 'smooth' regimes has been quantitatively described in Ref.~\cite{koshelev2020suppression} and results are summarized in Appendix \ref{App:Theory}
The correction to $\rho_s$ has two main temperature dependencies: via the ratio $T/\Delta_{0}(T)$, and via the correlation length $\xi_{h}(T)$. In the vicinity of $T_{m}$ the second of these dependencies is expected to dominate, whereas across a wider temperature range both are expected to contribute.

We have used SHM to image discrete vortices in high-quality RbEuFe\textsubscript{4}As\textsubscript{4} crystals, and studied the influence of the emerging magnetic order on the penetration depth, $\lambda(T)$, and the superfluid density, $\rho_{s}(T)$ \cite{bending1999local}. SHM has the advantage of being a quantitative, and non-invasive magnetic imaging technique that allows the magnetic penetration depth to be directly obtained from model fits. The temperature-dependent superfluid density has then been calculated assuming $\rho_{s}(T)\! \propto\! \lambda(T)^{-2}$, and exhibits a very substantial drop in the vicinity of T\textsubscript{m}. 
A direct comparison between our data and the 
model suggests that there must be a 
noticeable exchange interaction between the Eu$^{2+}$ moments and Cooper pairs that substantially suppresses superconductivity near T\textsubscript{m}. A recovery of the superfluid density at lower temperatures reflects a reduction of the magnetic correlation length 
and resulting weakening of the magnetic scattering.
The good qualitative agreement with our 
model represents an important step forward in our 
understanding of the subtle physics at play in magnetic superconductors.

\begin{figure}[htp]
\includegraphics[width=3.2in]{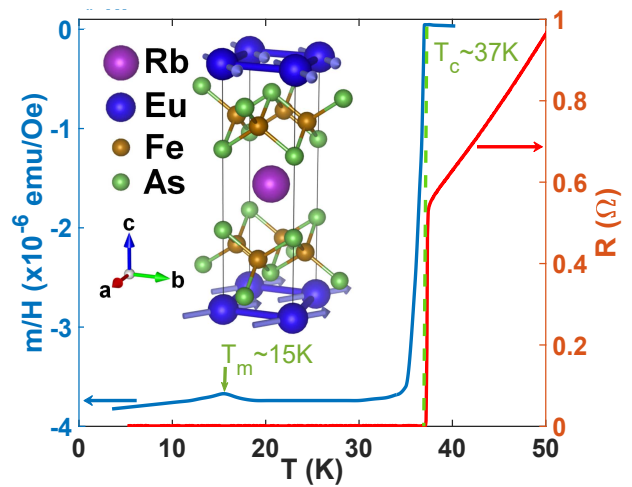}
\caption{\label{fig:1}Temperature dependence of the in-plane resistivity of RbEuFe\textsubscript{4}As\textsubscript{4} near the superconducting transition, and the magnetic susceptibility of a zero-field-cooled RbEuFe\textsubscript{4}As\textsubscript{4} single crystal with a 10Oe magnetic field applied along the c-axis. The inset shows one unit cell of RbEuFe\textsubscript{4}As\textsubscript{4}, where the magnetic structure of the Eu sublattice is indicated.}
\end{figure}

High-quality single crystals of RbEuFe\textsubscript{4}As\textsubscript{4} were grown using a RbAs flux, yielding flat, rectangular platelet-like crystals with lateral dimensions up to $\sim$1 mm in the \textit{ab} plane and thickness $\sim$60$\mu$m parallel to the \textit{c} axis \cite{bao2018single}. X-ray diffraction and specific heat measurements have previously confirmed that the crystals are single-phase material without EuFe\textsubscript{2}As\textsubscript{2} inclusions \cite{bao2018single}. RbEuFe\textsubscript{4}As\textsubscript{4} has a simple tetragonal structure and a P4/mmm space group, with one formula per unit cell and lattice constants $a = b = 3.88$\si{\angstrom} and $c = 13.27$\si{\angstrom} \cite{bao2018single}. A single unit cell of the crystal structure and atom-to-atom bonding is shown in the inset of Fig.\ \ref{fig:1}. The high quality of the crystals was confirmed via electronic transport and magnetization data.
The transport measurements were performed by attaching gold wires with silver paint in a standard 4-lead Hall bar configuration, and the in-plane 
resistivity then measured as a function of temperature. This is shown in Fig.\ \ref{fig:1}, revealing a superconducting transition of $\sim$37K. The magnetic susceptibility as a function of temperature was measured with a commercial magnetic property measurement system (MPMS3, Quantum Design) with magnetic fields applied along the c-axis revealing a magnetic transition at $\sim$15K.

\begin{figure}[htp]
\includegraphics{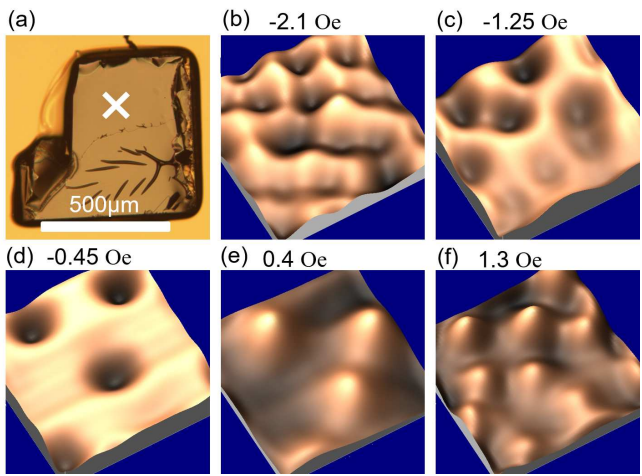}
\caption{\label{fig:2} (a) Optical micrograph of a RbEuFe\textsubscript{4}As\textsubscript{4} single crystal after cleaving and deposition of a conductive coating. The approximate location of the images taken in Fig. \ref{fig:3} is marked by the white cross. (b) - (f): Three dimensional SHM images of vortices in a RbEuFe\textsubscript{4}As\textsubscript{4} single crystal after field-cooling to 30K in effective perpendicular fields between -2.1Oe and 1.3Oe \cite{horcas2007wsxm}. The scan size is 12.6$\mu$m $\times$ 12.6$\mu$m. The full magnetic field range of the images (vertical scale) span 0.4G (-2.1Oe), 0.5G (-1.25Oe), 0.8G (-0.45Oe), 0.7G (0.4Oe), 0.5G (1.3Oe).}
\end{figure}

To prepare samples for SHM a crystal of RbEuFe\textsubscript{4}As\textsubscript{4} was glued flat on a gold-coated Si substrate and mechanically cleaved immediately prior to coating with a Cr(5nm)/Au(40nm) film (c.f., Fig.\ \ref{fig:2} (a)). This ensured good electrical contact between the scanning tunnelling microscopy (STM) tunnelling tip on the SHM sensor and the sample surface. The Hall probe used was based on a GaAs/AlGaAs heterostructure two-dimensional electron gas defined by the intersection of two 700nm wide wires. This was located $\sim$5$\mu$m from the gold-coated corner of a deep mesa etch acting as the STM tip \cite{bending1999local}. The Hall probe was mounted at an angle of approximately 1$^{\circ}$ with respect to the sample plane, ensuring that the STM tip is always the closest point to the sample surface. The Hall probe was approached to the sample until a threshold tunnel current was reached at which point the probe was manually lifted out of tunnelling by $\sim$50nm for rapid ‘flying mode’ scanning. From this a two-dimensional map of the magnetic induction across the surface of the sample was obtained \cite{bending1999local}, and several images were then averaged frame-by-frame to suppress low-frequency noise from the Hall probe.

Figure \ref{fig:2} (b) - (f) displays vortex-resolved SHM images for a RbEuFe\textsubscript{4}As\textsubscript{4} crystal after field-cooling to 30K from above T\textsubscript{c} at various small, perpendicular effective magnetic fields between -2.1Oe and 1.3Oe. Note that quoted values of magnetic field are effective ones after we have accounted for small amounts of flux that get trapped in our superconducting magnet upon initial cool down. This remanent field has been estimated by counting the number of vortices in our field-of-view at various applied fields, see Appendix \ref{App:EffField}. The scan range of the piezoelectric scanner is strongly temperature dependent and varies from 8.5$\mu$m $\times$ 8.5$\mu$m to 13.5$\mu$m $\times$ 13.5$\mu$m between 10K and 35K. 
Even below T\textsubscript{m}, 
we can attribute all the magnetic contrast in the images to vortices and see no sign of c-axis fields associated with domain walls between magnetic domains.
This differs from the MFM images in Ref.\ \cite{stolyarov2018unique} which showed the presence of such stray magnetic fields at temperatures below T\textsubscript{m}.
It is possible that these domain-wall fields are also present in our sample on much larger length scales than we probe in our measurements.

\begin{figure}[htp]
\includegraphics{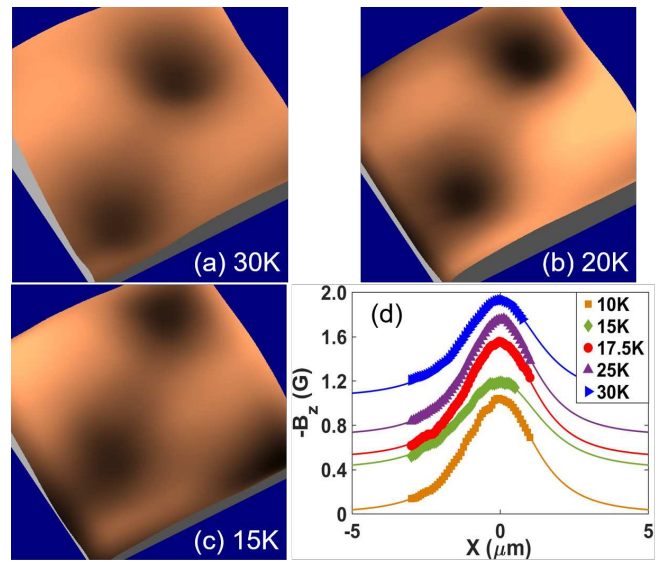}
\caption{\label{fig:3} 
Panels (a) - (c) display vortex-resolved SHM images after field-cooling in  $H_{z}^{eff}\! =\! -0.8$Oe from above T\textsubscript{c} to three different temperatures, illustrating the evolution of vortex profiles as T\textsubscript{m} is approached from above. The field of view in each of these images is 6.5$\mu$m $\times$ 6.5$\mu$m and vertical scales span 0.9G. (d) Vortex profiles extracted from SHM images in the sequence shown in (a)-(c) with superimposed fits to a modified Clem model, Eq.\ \eqref{clem}.}
\end{figure}

A sample was then field-cooled at $H_{z}^{eff}\! =\! -0.8$Oe from the normal state and images captured at several fixed temperatures down to 10K (c.f., Figs.\ \ref{fig:3} (a)-(c)). Profiles of one particular vortex at a few selected temperatures are presented in Fig.\ \ref{fig:3}(d). The influence of the long-range magnetic ordering is clearly reflected in the peak amplitude of the vortex which weakens (and broadens) as we approach 15K from above. The amplitude then starts to grow again at lower temperatures. The same behavior is observed in our detailed analysis of the temperature dependence of four distinct vortices in two different crystals. We also observe an unexpected increase in low-frequency noise in our images between 20K and 15K
in a regime where the intrinsic Hall sensor noise would normally fall as the temperature is lowered, 
see Appendix \ref{App:noise} 
for more details.
We tentatively associate this additional noise with magnetic fluctuations near the sample surface that have not been screened out by superconductivity. We also checked that there was no detectable hysteresis in the influence of the long-range magnetic order on the vortices by capturing images at both increasing and decreasing temperatures.\\

To investigate this behavior further, we have performed a quantitative analysis of the temperature-dependent vortex profiles $B_{z}(x_0,z,\lambda)$ by fitting them to a modified Clem model \cite{clem1975simple, kirtley2007upper} to extract the magnetic penetration depth, $\lambda(T)$,
\begin{widetext}
\begin{equation}
B_{z}(x_0,z,\lambda) = \frac{\Phi_{0}}{2\pi w^{2} \lambda K_{1}\!\left(\frac{\xi_{s}}{\lambda}\right)}
\int\limits^{w\!/2}_{-w\!/2}\!dy\!\int\limits^{x_0+w\!/2}_{x_0-w\!/2}\!\!dx
\int\limits^{\infty}_{0}\!qdq\frac{K_{1}\!\left(\sqrt{q^{2}\!+\!\lambda^{-2}}\xi_{s}\right)\exp(-qz)
J_{0}\!\left(q\sqrt{x^{2}\!+\!y^{2}}\right)}{\sqrt{q^{2}+\lambda^{-2}}+q},\label{clem}
\end{equation}
\end{widetext}
where $z$ is the sensor scan height measured from the sample surface, $w = 0.5\mu$m is the electronic width of the Hall probe, $q$ is the Fourier wave vector, $K_{1}$ and $J_{0}$ are Bessel functions, 
$\xi_{s}$ is the coherence length for which we assume $\xi_{s}\!=\!1.46$nm$/\sqrt{1\!-\!T/T_{c}}$,
and $\Phi_{0}$ is the flux quantum. 
Fits to Eq.\ (\ref{clem}) have been superimposed on the measured profiles in Figure \ref{fig:3} (d) showing excellent agreement. The value of $z$ = 1.45$\pm$0.01$\mu$m was extracted from a fit at 30K with $\lambda(30K)$ estimated from experimental data using Ginzburg-Landau theory expressions for the specific heat jump and upper critical field slope at $T_{c}$ (c.f. supplementary materials). This is consistent with the sensor tilt angle used. The same scan height was maintained at all other temperatures and, although it is large compared to the penetration depth we are trying to measure, we are nevertheless able to extract values of $\lambda(T)$ from fits with good accuracy.
\begin{figure}[htp]
\includegraphics{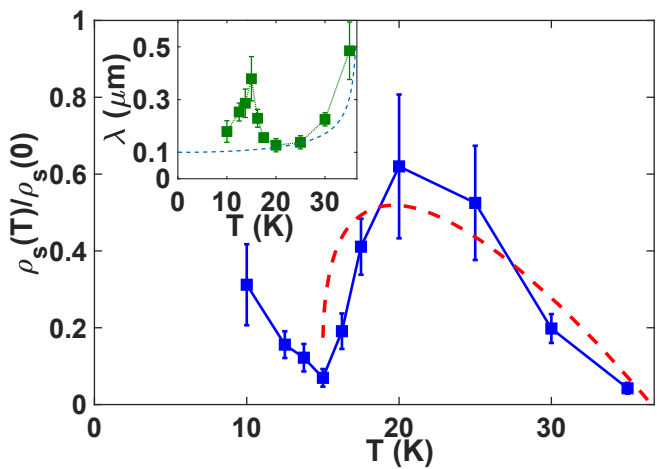}
\caption{Temperature dependence of the normalized superfluid density, $\rho_{s}(T)/\rho_{s}(0)$ (solid symbols), and a fit to the model described in the text (dashed line). The inset shows a plot of the penetration depth as a function of temperature extracted from fits to a modified Clem model, Eq.\ \eqref{clem}, for one of the vortices shown in Figs.\ \ref{fig:3} (a)-(c). 
The dashed light blue line shows the assumed bare penetration depth dependence, $\lambda_0(T) = \lambda(0)/\sqrt{1-(T/T_{c})^{2}}$.
\label{fig:4}}
\end{figure}
The temperature dependence of the extracted London penetration depth $\lambda(T)$ is shown in the inset of Fig.\ \ref{fig:4}, while the corresponding normalized superfluid density, $\rho_{s}(T)/\rho_{s}(0) = \lambda(0)^{2}/\lambda(T)^{2}$, is plotted in the main panel of this figure.
The vertical error bars on experimental data points reflect the impact of the sensor noise level on the vortex profile fitting process combined with uncertainties in the estimated scan height, $z$.

The most natural mechanism of the observed significant enhancement of the vortex magnetic size near $T_m$ is suppression of superconductivity due to exchange interaction between Cooper pairs and localized Eu$^{2+}$ moments.
This enhancement 
becomes especially pronounced in the vicinity of the magnetic transition, where the moments become strongly correlated. The quantitative description of the suppression of superconducting parameters by correlated magnetic fluctuations has been elaborated in Ref.\ \cite{koshelev2020suppression}. In Appendix \ref{App:Theory}, 
we summarize the results for the correction to the superfluid density which we use for the modelling of the data. The relative correction is proportional to the square of the amplitude of the exchange field, $h_0$, and depends on two ratios, $T/\Delta_0(T)$ and $\xi_s(T)/\xi_h(T)$.  
We also account for renormalization of parameters due to the nonlocality of the exchange interaction described by the range $a_J$, $\xi_h^2=\xi_S^2+2a_J^2$ and $\tilde{h}_0^2=h_0^2\xi_S^2/\xi_h^2$, where $\xi_S$ is the spin correlation length. We assume the Berezinskii-Kosterlitz-Thouless (BKT) shape for the latter, $\xi_{S}(T)\!=\!a\exp[b\sqrt{T_{m}/(T\!-\!T_{m})}]$, based on recent experimental observations \cite{hemmida2020topological}, where  $a\! =\! 0.39$nm is the in-plane Eu atom spacing, and we use the numerical factor $b$ as a fit parameter.

We plot the results of our model as a dashed line alongside our data using the amplitude of the exchange field, $h_{0}\! =\! 15$K, a zero temperature superconducting gap, $\Delta (0)\! =\! 2$meV, a BKT constant $b\! =\! 1$, and the nonlocality range $a_J\! =\! 3$a.
We use the BCS temperature dependence to describe $\Delta$(T), which is corrected in the model for the magnetic exchange interaction. The Ginzburg-Landau coherence length is estimated to be $\xi^{GL}\! =\! 1.46$nm, deduced from the linear slope of the c-axis upper critical field near T\textsubscript{c} \cite{smylie2018anisotropic, willa2019strongly}.
The exchange correction is added to the bare penetration depth $\lambda_0(T)$ that would be realized in the absence of any exchange interactions with magnetic fluctuations, for which we assume a simple phenomenological temperature dependence $\lambda_0(T) = \lambda(0)/\sqrt{1-(T/T_{c})^{2}}$. We apply a value of $\lambda(0)\! =\! 100$,  which has been estimated from fitting the aforementioned dependence to our extracted penetration depths from 20K and above. This value provides a better description across all temperatures above 20K than 
the value $\lambda(0)\! =\! 133$nm extracted from the thermodynamic data using the Ginzburg-Landau theory, see Appendix \ref{App:LambEst}.

The strong suppression of superfluid density in the vicinity of $T_{m}$ is remarkable and was not previously observed in RbEuFe\textsubscript{4}As\textsubscript{4} or indeed any other ferromagnetic-superconductor, although this possibility was recently suggested by Willa \emph{et al.} \cite{willa2019strongly}. This is also in apparent conflict with the analysis of recent ARPES measurements which concluded that the two sublattices are almost fully decoupled \cite{kim2020superconductivity}. To understand the temperature-dependent trend of the suppression seen in our data in Fig. \ref{fig:4}, we turn to our model. Comparing the measured normalized $\rho_{s}$ with predictions of the model, we find good qualitative agreement with our 2D BKT description of the magnetic correlations above the magnetic transition temperature, T\textsubscript{m}\! =\! 15K, confirming the nature of the ordering and its impact through $\xi_{h}(T)$ on superconductivity in the vicinity of T\textsubscript{m}. The temperature-dependent magnetic correlation length, $\xi_{h}(T)$, which is governed by the constant, $b$, is responsible for the wide temperature range over which the magnetic ordering influences the superconducting parameters. Above T\textsubscript{m}, the suppression decreases rapidly as temperature increases, while the shift in $T_{c}$ is $\sim$ 1K. Although the suppression of superfluid density is quite large, the fitted magnetic exchange constant remains moderate at $h_{0}\!\approx \! 0.4T_{c}$. This is still significantly smaller than exchange constants estimated for the ternary compounds, which are several orders of magnitude larger than T\textsubscript{c} \cite{Kulic2008InBook}. This suggests a weak enough coupling between Eu moments and Cooper pairs in our material that superconductivity is never destroyed, yet one that is strong enough to have a substantial impact on the superconducting parameters near T\textsubscript{m}. We emphasize that our model is a qualitative one and at best only qualitative agreement with our data is expected. In particular, the model assumes two-dimensional scattering 
behavior across the whole temperature range, $T>T_{m}$, while this assumption must break down in the vicinity of the magnetic transition where 
a crossover to a three-dimensional regime takes place.
In addition, the BCS expressions we have used for the temperature dependence of the gap and penetration depth were derived for single-band materials, whereas RbEuFe\textsubscript{4}As\textsubscript{4} has a more complicated multiband structure. Nevertheless, the qualitative agreement between the model and data validates our simple assumptions in this fascinating magnetic superconducting material.

In conclusion, we have directly quantified the temperature-dependent superfluid density in RbEuFe\textsubscript{4}As\textsubscript{4} crystals to reveal a significant suppression of superconductivity due to correlated quasi-two-dimensional magnetic fluctuations, despite the apparent spatial separation of the two sublattices. Although insufficient to completely destroy superconductivity, this suggests a significant influence of the exchange interaction on the superconducting subsystem. Our results will stimulate additional investigations into the properties of RbEuFe\textsubscript{4}As\textsubscript{4}, and other magnetic-superconductors, building on the existing analytical model. 

We acknowledge support from the U.S. Department of Energy, Office of Science, Basic Energy Sciences, Materials Sciences and Engineering Division for the crystal growth, theoretical modelling and magnetotransport measurements. Financial support was also provided by the Engineering and Physical Sciences Research Council (EPSRC) in the UK under grant nos. EP/R007160/1, and the Nanocohybri COST Action CA-16218. D.C. acknowledges financial support from the Lloyds Register Foundation ICON (award nos. G0086) and L.F. from the EPSRC Centre for Doctoral Training in Condensed Matter Physics, Grant No. EP/L015544/1.
\newpage
\appendix

\section{Theoretical model for suppression of superfluid density by correlated
	quasi-two-dimensional magnetic fluctuations}
\label{App:Theory}

The suppression of the superconducting gap and superfluid density
by correlated magnetic fluctuations in the vicinity of the magnetic
transition has been evaluated recently in Ref.~\citep{koshelev2020suppression}
and here we summarize the results for the superfluid density with
small generalization accounting for the nonlocality of the exchange
interactions. The models assumes a clean layered magnetic superconductor
in which a continuous magnetic transition takes place inside superconducting
state. Increase of the exchange-field correlation length $\xi_{h}$
in the vicinity of the magnetic transition enhances suppression of
superconductivity. The influence of nonuniform exchange field on superconducting
parameters is very sensitive to the relation between $\xi_{h}$, and
superconducting coherence length $\xi_{s}$ defining the 'scattering'
($\xi_{h}<\xi_{s}$) and 'smooth' ($\xi_{h}>\xi_{s}$) regimes. The
model in Ref.~\citep{koshelev2020suppression} provides a quantitative description
of this 'scattering-to-smooth' crossover for the case of quasi-two-dimensional
magnetic fluctuations. 

For a material composed of magnetic and superconducting layers, the
exchange field acting on the conduction electron spins has the form
\begin{equation}
	\boldsymbol{h}_{n}(\boldsymbol{r})\!=\!-\!\sum_{m,\boldsymbol{R}}J_{nm}(\boldsymbol{r}\!-\!\boldsymbol{R})\boldsymbol{S}_{m}(\boldsymbol{R})\label{eq:ExField}
\end{equation}
where the indices $n$ and $m$ correspond to conducting and magnetic
layers, respectively, $\boldsymbol{S}_{m}(\boldsymbol{R})$ are localized
spins, and $J_{nm}(\boldsymbol{r}\!-\!\boldsymbol{R})$ are the exchange
constants. In the paramagnetic state the field $\boldsymbol{h}_{n}(\boldsymbol{r})$
is random and the Fourier transform of the exchange-field correlation
function for an isolated layer is
\begin{equation}
	\left\langle \left|\boldsymbol{h}_{\mathbf{\mathbf{q}}}\right|^{2}\right\rangle =\sum_{m}\left[J_{nm}(\boldsymbol{q})\right]^{2}\left\langle \left|\boldsymbol{S}_{\mathbf{\mathbf{q}}}\right|^{2}\right\rangle .\label{eq:hqSq}
\end{equation}
We neglect magnetic correlation between different magnetic layers
and assume the spin correlation function for an isolated layer in
the standard form
\begin{equation}
	\left\langle \left|\boldsymbol{S}_{\mathbf{\mathbf{q}}}\right|^{2}\right\rangle =\frac{2\pi S_{0}^{2}\xi_{S}^{2}/\ln\left(\xi_{S}/a\right)}{1+\xi_{S}^{2}q^{2}}.\label{eq:Sq}
\end{equation}
It is normalized by the condition $\int\frac{d^{2}\boldsymbol{q}}{(2\pi)^{2}}\left\langle \left|\boldsymbol{S}_{\mathbf{\mathbf{q}}}\right|^{2}\right\rangle =S_{0}^{2}$,
where $S_{0}$ is the magnitude of spin ($7/2$ for Eu$^{2+}$). Here $\xi_{S}$ is the spin correlation length diverging at the magnetic transition. The spin correlation function in real space is $\langle S(0)S(r)\rangle=S_0^2\exp(-r/\xi_S)$. As
in RbEuFe$_{4}$As$_{4}$ the magnetic and superconducting layers
are separated, the exchange interaction is most likely is indirect
and has significant nonlocality. We assume a simple shape for $J_{nm}(\boldsymbol{q})$,
\begin{equation}
	J_{nm}(\boldsymbol{q})=\frac{\mathcal{J}_{nm}}{1+a_{J}^{2}q^{2}},\label{eq:Jnmq}
\end{equation}
where $a_{J}$ is the nonlocality range, which is probably $2-3$
lattice spacings. Then we have
\[
\left\langle \left|\boldsymbol{h}_{\mathbf{\mathbf{q}}}\right|^{2}\right\rangle =\sum_{m}\mathcal{J}_{nm}^{2}\frac{2\pi S_{0}^{2}\xi_{S}^{2}/\ln\left(\xi_{S}/a\right)}{\left(1\!+\!a_{J}^{2}q^{2}\right)^{2}\left(1\!+\!\xi_{S}^{2}q^{2}\right)}.
\]
To capture the qualitative effect of crossover at $\xi_{S}\sim a_{J}$,
we approximate this correlation function with
\[
\left\langle \left|\boldsymbol{h}_{\mathbf{\mathbf{q}}}\right|^{2}\right\rangle \approx\frac{2\pi h_{0}^{2}\xi_{S}^{2}/\ln\left(\xi_{S}/a\right)}{1+\left(2a_{J}^{2}+\xi_{S}^{2}\right)q^{2}}
\]
with $h_{0}^{2}=\sum_{m}\mathcal{J}_{nm}^{2}S_{0}^{2}$ is the maximum
amplitude of the exchange field. The last result can be rewritten
as 
\begin{equation}
	\left\langle \left|\boldsymbol{h}_{\mathbf{\mathbf{q}}}\right|^{2}\right\rangle =\frac{2\pi\tilde{h}_{0}^{2}\xi_{h}^{2}/\ln\left(\xi_{S}/a\right)}{1+\xi_{h}^{2}q^{2}},\label{eq:hq-Renorm}
\end{equation}
where $\xi_{h}^{2}=2a_{J}^{2}+\xi_{S}^{2}$ and $\tilde{h}_{0}^{2}=h_{0}^{2}\frac{\xi_{S}^{2}}{2a_{J}^{2}+\xi_{S}^{2}}$
are the correlation length and exchange-field amplitude renormalized by the nonlocality of the exchange interaction. Therefore nonlocality increases the effective correlation length and reduces the effective amplitude of the exchange field. This effect becomes noticeable when temperature is not too close the to magnetic transition when the spin correlation length becomes comparable with the nonlocality range.

The correction of the to $\lambda^{-2}$ caused by a nonuniform exchange field with the correlation function in Eq.~\eqref{eq:hq-Renorm} evaluated in Ref.~\citep{koshelev2020suppression} is given by
\begin{widetext}
\begin{align}
	\lambda_{1}^{-2}(T) & \!=\!-\lambda_{0}^{-2}(T)\frac{\tilde{h}_{0}^{2}}{2\Delta_{0}^{2}(T)\ln\left(\xi_{S}(T)/a\right)}\mathcal{V}_{Q}\left(\frac{2\pi T}{\Delta_{0}(T)},\frac{\xi_{s}(T)}{\xi_{h}(T)}\right),\label{eq:LambCorrCorrFl}
\end{align}
where $\lambda_{0}(T)$ and $\Delta_{0}(T)$ are the unperturbed
values of the London penetration depth and the gap, $\xi_{s}(T)=v_{F}/2\Delta_{0}(T)$
is the coherence length. The reduced function $\mathcal{V}_{Q}\left(\tilde{T},\alpha_{h}\right)$
is defined by relations
\begin{align}
	\mathcal{V}_{Q}\left(\tilde{T},\alpha_{h}\right)\! & =\!\left[\mathcal{D}\left(\tilde{T}\right)\right]^{-1}\!\tilde{T}\sum_{n=0}^{\infty}\left\{ K_{Q}\left[\tilde{T}(n\!+\tfrac{1}{2})\right]\mathcal{V}_{\Delta}\!\left(\tilde{T},\alpha_{h}\right)\!+\!R_{Q}\left[\tilde{T}(n\!+\tfrac{1}{2}),\alpha_{h}\right]\right\} ,\label{eq:VQ}\\
	\mathcal{D}\left(\tilde{T}\right)\! & =\!\tilde{T}\sum_{n=0}^{\infty}\left(\left[\tilde{T}(n\!+\tfrac{1}{2})\right]^{2}\!+\!1\right)^{-3/2},\label{eq:KQ}
\end{align}
where we used the reduced variables $\tilde{T}=2\pi T/\Delta_{0}(T)$
and $\alpha_{h}=\xi_{s}(T)/\xi_{h}(T)$. Here the first term in the
curly brackets in Eq.~\eqref{eq:VQ} is due to the gap reduction
by magnetic scattering and the functions $K_{Q}\left(z\right)$ and
$\mathcal{V}_{\Delta}\left(\tilde{T},\alpha_{h}\right)$ are defined
as
\begin{align}
	K_{Q}\left(z\right) & \!=-\frac{\partial}{\partial z}\frac{z}{\left(z^{2}\!+\!1\right)^{3/2}},\label{eq:VD-T-a}\\
	\mathcal{V}_{\Delta}\left(\tilde{T},\alpha_{h}\right) & \!=\left[\mathcal{D}\left(\tilde{T}\right)\right]^{-1}\!\tilde{T}\sum_{n=0}^{\infty}R\left[\tilde{T}(n\!+\!\tfrac{1}{2}),\alpha_{h}\right],
\end{align}
with 
\begin{align}
	R\left(z,\alpha_{h}\right)\! & =\frac{1}{\left(z^{2}\!+\!1\right)^{3/2}\left(z^{2}\!+\!1\!-\!\alpha_{h}^{2}\right)}\left[1\!+\!\left(2z^{2}\!-\!1\!-\!\frac{2z^{2}\alpha_{h}^{2}}{z^{2}\!+\!1}\right)\mathit{L}\left(z,\alpha_{h}\right)\right],\\
	\mathit{L}\left(z,\alpha_{h}\right) & \!=\begin{cases}
		\frac{\sqrt{z^{2}+1}}{\sqrt{z^{2}+1-\alpha_{h}^{2}}}\ln\left(\frac{\sqrt{z^{2}\!+\!1}\!+\!\sqrt{z^{2}\!+\!1\!-\!\alpha_{h}^{2}}}{\alpha_{h}}\right), & z^{2}\!>\!\alpha_{h}^{2}\!-\!1\\
		\frac{\sqrt{z^{2}+1}}{\sqrt{\alpha_{h}^{2}-z^{2}-1}}\arctan\frac{\sqrt{\alpha_{h}^{2}\!-\!z^{2}\!-\!1}}{\sqrt{z^{2}\!+\!1}}, & z^{2}\!<\!\alpha_{h}^{2}\!-\!1
	\end{cases}.
\end{align}
\end{widetext}
The function $R_{Q}\left(z,\alpha_{h}\right)$ in Eq.~\eqref{eq:VQ}
describes the direct influence of the magnetic scattering on the superfluid density and is defined as 
\begin{equation}
	R_{Q}\left(z,\alpha_{h}\right)=\frac{\partial}{\partial z}\frac{z\left[1-\left(3-\frac{2\alpha_{h}^{2}}{z^{2}\!+\!1}\right)L(z,\alpha_{h})\right]}{\left(z^{2}\!+\!1\right)^{3/2}\left(z^{2}\!+\!1\!-\!\alpha_{h}^{2}\right)}.\label{eq:RQ}
\end{equation}
We use Eq.~\eqref{eq:LambCorrCorrFl} to model the experimental behavior of $\lambda^{-2}(T)$. The necessary input model parameters are the temperature dependent gap $\Delta_{0}(T)$ which also determines the coherence length $\xi_{s}(T)$, spin correlation length $\xi_{S}(T)$, the bare strength of exchange
field $h_{0}$, and the nonlocality range $a_{J}$. We assumed the
Berezinskii-Kosterlitz-Thouless shape for the magnetic length,  $\xi_{S}(T)\!=\!a\exp[b\sqrt{T_{m}/(T\!-\!T_{m})}]$ and treated the nonuniversal numerical constant $b$ as an additional fit parameter.

\section{Selection of model parameters}

\begin{figure}[htp]
	\includegraphics{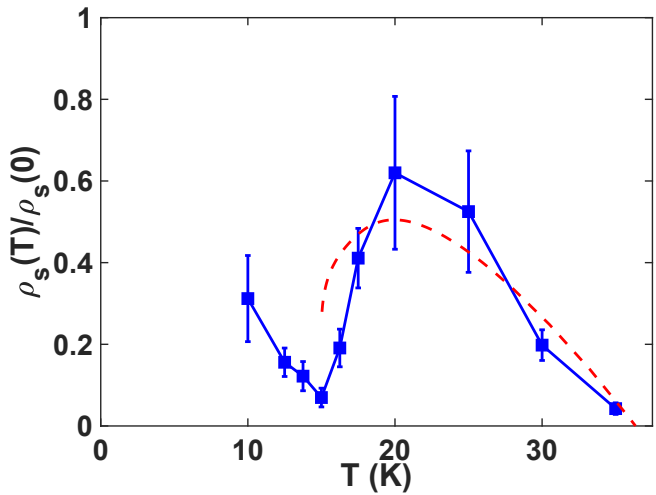}
	\caption{\label{fig:S2} 
		Temperature dependence of the normalised superfluid density, $\rho_{s}(T)/\rho_{s}(0)$ (solid symbols), as in Fig.\ 4 of the main text, and a fit to the model with BCS value of the gap, $\Delta (0)\! =\! 5$meV, exchange field $h_{0}\! =\! 50$K, a BKT constant $b\! =\! 1$, and the nonlocality range $a_J\! =\! 4$a (dashed line).}
\end{figure}
Since RbEuFe\textsubscript{4}As\textsubscript{4}, as other iron-pnictide superconductors, has multiple bands with different superconducting gaps, our model is only capable to provide qualitative description of the data. To reduce uncertainties of the model, we assume the BCS temperature dependence of the gap $\Delta_{0}(T)$ but leave the zero-temperature gap $\Delta_{0}(0)$ as a free parameter.  Other three parameters of the model are the bare strength of the exchange field $h_{0}$, the constant $b$ in the BKT temperature dependence of the spin correlation length, and the nonlocality range $a_J$. On general grounds, as we observe a substantial suppression of the superfluid density near $T_m$, the model requires $h_{0}$ comparable with  $\Delta_{0}(0)$. The shape of the temperature dependence is sensitive to the parameters $b$ and $a_J$. 

If we assume that the zero-temperature gap $\Delta_{0}(0)$ has the BCS value $\approx 5$meV, which also coincides with the value extracted from the optical data \cite{stolyarov2018unique}, then the model describes our data if we take $h_{0}\! =\! 50$K, $b\! =\! 1$, and $a_J\! =\! 4$a, see Fig.\  \ref{fig:S2}. However, this value of $h_{0}$ looks unrealistically high  as it exceeds the Eu to Eu moment interaction yielding the magnetic transition temperature, 15K. Also, the nonlocality range is somewhat higher than expected. This is why we assumed the smaller value of  $\Delta_{0}(0)=2$meV, for which the data can be modeled with more reasonable values of  $h_{0}\! =\! 15$K and $a_J\! =\! 3$a. The value $b=1$  does not change. The model curve with these parameter is shown in Fig.\ 4 of the main text and it provides a somewhat better description of our data in comparison with the first set.

\section{Estimation of $\lambda(0)$ from the thermodynamic data}
\label{App:LambEst}

The value of the bare penetration depth at zero temperature, $\lambda (0)$, used to evaluate the normalized superfluid density has been calculated from experimental data measured at high temperatures close to T$_{c}$, where the influence of magnetic fluctuations is very weak. To achieve this, the results of Ginzburg-Landau (GL) theory have been used to estimate the GL value $\lambda_{\mathrm{GL}}$ directly from the specific heat jump   
per unit volume, 
$\Delta C_V/T_{c} = 1.7\cdot 10^{-3}$ J/cm$^{3}$K$^{2}$
and the slope of $H_{c2}(T)$, 
$dH_{c2}/dT = -4.2$ T/K,
at $T_{c}$ \cite{smylie2018anisotropic,willa2019strongly}
into the following equation %
(in CGS units),
\begin{equation}
	\lambda_{\mathrm{GL}}^{2}=\frac{\Phi_{0}}{16\pi^{2}\Delta C_{V}}\left|\frac{dH_{c2}}{dT}\right|.
	\label{eq:LAMBDA}
\end{equation}
Substitution of parameters gives $\lambda_{\mathrm{GL}}=94$nm. For the assumed temperature dependence $\lambda(T)=\lambda(0)/\sqrt{1-(T/T_c)^2}$, this gives the zero temperature value $\lambda(0)=\sqrt{2}\lambda_{\mathrm{GL}}\approx 133$nm.

\section{Magnetic noise fluctuations above the magnetic transition temperature}
\label{App:noise}

\begin{figure}[htp]
	\includegraphics{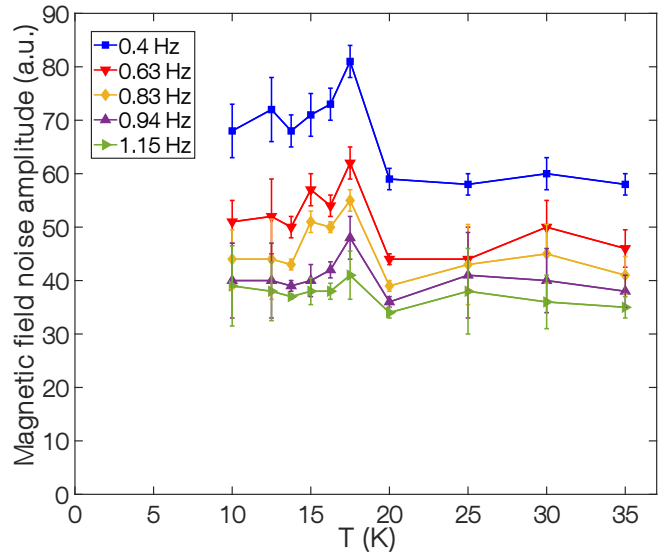}
	\caption{\label{fig:S1} 
		Magnetic field noise amplitude extracted from 2D FFT linescans at various frequencies as a function of temperature. As the temperature is reduced the low frequency noise rises to a peak around 17.5K, before falling again as the temperature is lowered further down to 10K.}
\end{figure}

While performing magnetic imaging of RbEuFe\textsubscript{4}As\textsubscript{4}, we noticed an unexpected increase in very low frequency Hall sensor noise in images captured between 20K and 15K. Sensor noise would normally reduce at lower temperatures due to the lower thermal noise contribution to the Hall probe signal. We tentatively associate the observed increase with the detection of magnetic fluctuations near the sample surface. To investigate this further we have generated 2D FFT spectra of the images to estimate the 'spatial' noise amplitude (which can be directly related to temporal noise amplitude) at various frequencies and temperatures, as plotted in Fig. \ref{fig:S1}. To generate Fig. \ref{fig:S1} we take a linescan across a 2D FFT parallel to the slow scan direction and fit this to a Lorentzian profile. Since all images take approximately 240 seconds to complete, this allows us to directly convert between spatial frequencies and temporal frequencies.

\section{Effective fields determined by vortex count}
\label{App:EffField}

The effective fields stated in the main manuscript are defined as the remnant field estimated by counting the number of vortices in the field-of-view, then added to the externally applied field. In Fig \ref{fig:S2} we show the plot of the vortex number vs applied field to show the robustness of this protocol in the cooldown cycles for the images captured in Fig. 2 in the main manuscript.

\begin{figure}[htp]
	\includegraphics{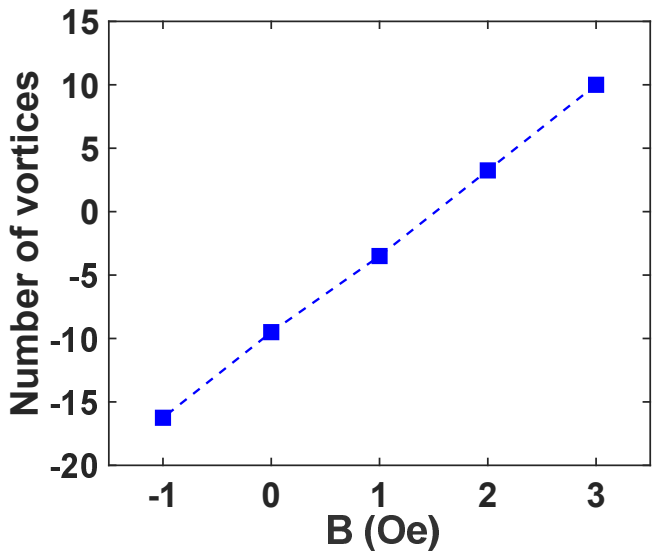}
	\caption{\label{fig:S2} Vortex number as a function of the applied perpendicular field, showing the expected linear relationship. The offset is used to determine the effective fields stated in the main text.}
\end{figure}



\newpage
\bibliography{VortImageRbEu1144}

\end{document}